# BEAM DYNAMICS AND WAKE-FIELD SIMULATIONS FOR HIGH GRADIENT ILC LINACS

C.J. Glasman & R.M. Jones; Cockcroft Institute, Daresbury, WA4 4AD, UK; University of Manchester, Manchester M13 9PL, UK.

*Abstract*

Higher order modes (HOMs) are simulated with finite element and finite difference computer codes for the ILC superconducting cavities. In particular, HOMs in KEK's Ichiro type of cavity and Cornell University's Reentrant design are focused on in this work. The aim, at these universities and laboratories, is to achieve an accelerating gradient in excess of 50 MV/m in 9-cell superconducting cavities whilst maintaining a high quality and stable electron beam. At these gradients, electrical breakdown is an important cause for concern and the wakefields excited by the energetic electron beams are also potentially damaging to the beam's emittance. Here we restrict the analysis to performing detailed simulations, on emittance dilution due to beams initially injected with realistic offsets from the electrical centre of the cavities. We take advantage of the latest beam dynamics codes in order to perform these simulations.

## INTRODUCTION

The main superconducting (SC) linacs of the ILC will accelerate electron (positron) beams from energies of 5 (15) GeV to a center of mass energy of 500 GeV at collision. Efficient operation of the machine demands high luminosity collisions which are achieved by accelerating a train of 2625 low emittance bunches particles. This low emittance must of course be preserved in transport through the main linacs and beam delivery system to the interaction point.

Here we consider the main positron linac where the beam quality can be degraded and the emittance can be diluted due to a number of factors including energy spread, phase jitter in cavities, beam position feedback errors, quadrupole magnet misalignments and wakefield effects. We focus on the dilution in the transverse emittance due to long range transverse wakefields. In this case the wakefields are excited by the relativistic particle beam and consist of a series of higher order modes [1]. These wakefields have the potential to not only dilute the emittance, but can also cause a beam beak up instability (BBU) [2].

The ILC will employ superconducting cavities operating at gradients at 31.5 MV/m. It is envisaged that the cavity design will be the TESLA type which has been developed at DESY over a long period [3]. TESLA is a relatively mature design, however there are still significant concerns regarding the limited yield and reproducibility of high gradient cavities.

There is also a concerted international effort focused on increasing the gradient of the SC cavities. Reshaping the cavity has allowed the accelerating gradient to be increased without pushing the magnetic field past the quenching limit (~180 mT) on the walls of the cavity [4]. This has allowed cavity designs which in theory will sustain accelerating gradients in excess of 50 MV/m. There is also a recent design which minimizes both the electric field as well as the magnetic field on the walls of the cavity [5].

In practice only single-cell cavities have reached these gradients, at Cornell University [6] with the Reentrant design and at KEK with the Ichiro design [7]. Complete 9-cell cavities are in the process of being fabricated and tested for both designs with a view to achieving similar gradients. The new high gradient designs operate at the same accelerating frequency as the baseline TESLA design. Furthermore the cavity shapes are similar to that of the TESLA design.

However, the perturbations of the cavity geometry are expected to give rise to a modification of the mode distribution. With this in mind, we carefully investigated the kick factors and mode frequencies in the alternative designs. The transverse long range wakefield experienced by the bunch train is given by:

$$W_T(t) = 2\sum_p K_p \sin(\omega_p t) e^{\frac{-\omega_p t}{2Q_p}}$$

where $\omega_p/2\pi$, $K_p$ and $Q_p$ are the modal frequencies, kick factors [1] and damping Qs respectively for the dipole mode $p$.

These modes have been simulated in detail [8, 9] using parallel finite difference and finite element codes GdfidL [10] and Analyst [11] and this data has been used as input for beam dynamics simulations presented herein, using the codes Lucretia [12] and Placet [13]. The codes track particle bunches through the lattice of the linac in which the energy is increased from 15 GeV to 250 GeV.

The envelope of the transverse long range wakefield is displayed in Fig. 1 with a damping Q imposed on all modes of $10^5$ for the Ichiro cavity. Here it is evident that after 500 bunches (~50 km) the wake has decayed by 6 orders of magnitude. Thus, in all beam dynamics simulations 500 bunches is sufficient to account for the interaction between the wakefield and the multi-bunch train. We report on the results of these simulations in the next section.

## BEAM DYNAMICS SIMULATIONS

In constructing the main linacs of ILC approximately 16,000 cavities will be required. Fabrication of these cavities is expected to introduce random errors in the geometry and thereby alter the modal frequencies and kick factors for each cavity. These errors are in fact necessary in order that the beam receives a random kick from cavity to cavity. A linac made up of identical cavities would impart the same kick coherently on the beam from cavity to cavity. This would resonantly drive beam break up and lead to severe emittance dilution.

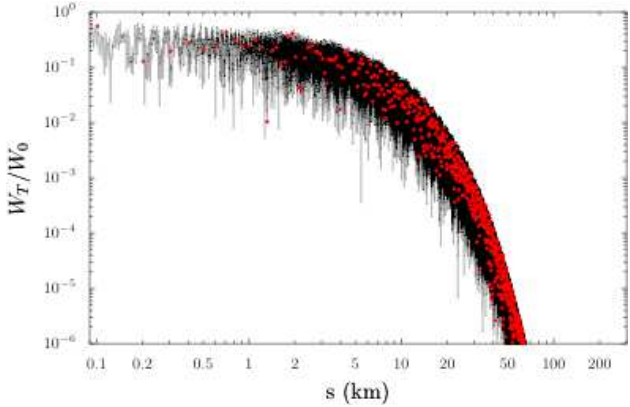

Figure 1: Envelope of long range transverse wakefield for the Ichiro cavity with $Q=10^5$, $W_0=0.1472$ V/pC/mm/m. Points show the location of the bunches.

The simulations described here are preformed with the Matlab-based code Lucretia. All simulations incorporate manufacturing errors by generating 100 sets of dipole mode frequencies in which each mode has been shifted by a random number generated according to a normal distribution; these define 100 different cavity types. These cavity types are then randomly distributed throughout the linac prior to tracking the beam down the linac. We applied a 1 MHz RMS spread in the dipole mode frequencies and a uniform damping Q of $10^6$. Each bunch was subjected to an injection offset of $\sigma_y$ (~ 6 μm) and trains of 500 bunches were subsequently tracked down the linac at the nominal bunch spacing of 369ns [14]. Placet simulations were also performed, although in this case the cavity mode randomization is achieved in a somewhat different manner; in this case all modes are shifted by the same random factor for each cavity in the linac. Fig.2 illustrates the dilution in the projected emittance for the bunch train at each beam position monitor position in the lattice that results from tracking down 40 machines. Each curve results from a different random seed, and the frequency distributions are distributed randomly over the complete linac. Figs. 3 and 4 present the results for machines consisting of Reentrant and Ichiro cavities, respectively. Fig. 5 displays a comparison of the mean value of the emittance dilution for each cavity type, together with the results of Placet simulations for the Ichiro cavity.

There is a discrepancy between the Lucretia and Placet results for the linac made up of Ichiro cavities. The difference between individual tracking simulations, for Lucretia, illustrated in Fig. 4, was small (~1%) for all cavity types simulated with Placet.

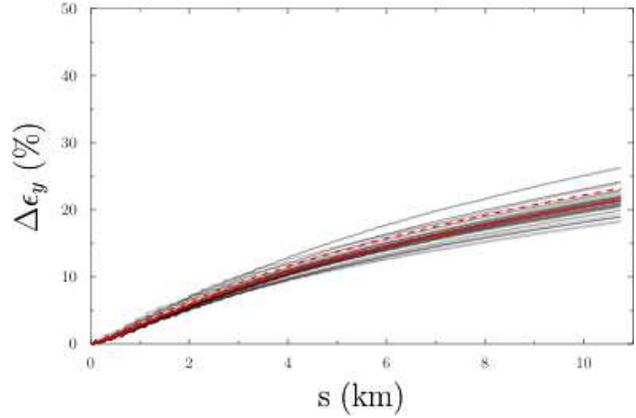

Figure 2: Emittance dilution simulated with Lucretia for 40 machines consisting of TESLA cavities. The mean dilution over all machines is displayed in red, with one standard deviation dashed.

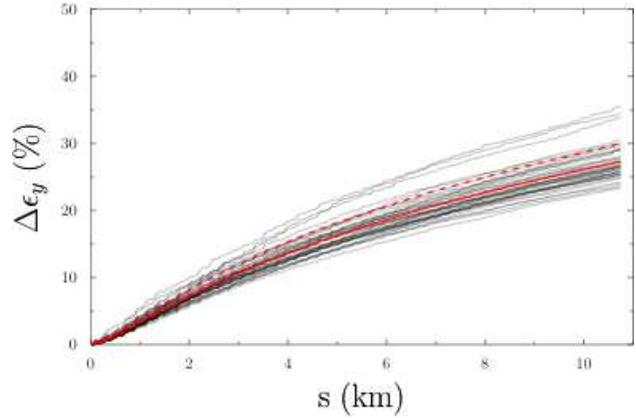

Figure 3: Emittance dilution simulated with Lucretia for 40 machines consisting of Reentrant cavities. The mean curve is displayed in red, with one standard deviation dashed.

The beam dynamics simulations described above incorporated random cavity fabrication errors but did not consider systematic effects.

In practice, superconducting cavities are squeezed to tune and to optimally achieve field flatness in the 1.3 GHz accelerating mode. This process changes the cell geometry and gives rise to a shift in the designed dipole mode frequencies. This tuning effect and those resulting from an amalgam of relaxed tolerances in the manufacturing process can readily make the dipole frequencies significantly detuned. This leads to both systematic and random frequency errors in the dipole modes. The sinusoidal dependence in Eq. 1 indicates that the wakefield time coordinate is determined by the bunch spacing. Consequently shifts in the bunch spacing are to a large extent equivalent to a shift in modal frequencies. Shifting the modal frequencies can push the beam dynamics into a region in which resonant BBU occurs. Clearly these regimes need to be predicted prior to fabrication of the complete linac.

The onset of this instability is predicted by the RMS of the sum wakefield which can be evaluated for small shifts in bunch spacing. Fig. 6 reveals a series of resonances associated with the Ichiro cavity. Applying heavier damping, changing the Qs from $10^6$ to $10^5$, suppresses these resonances significantly.

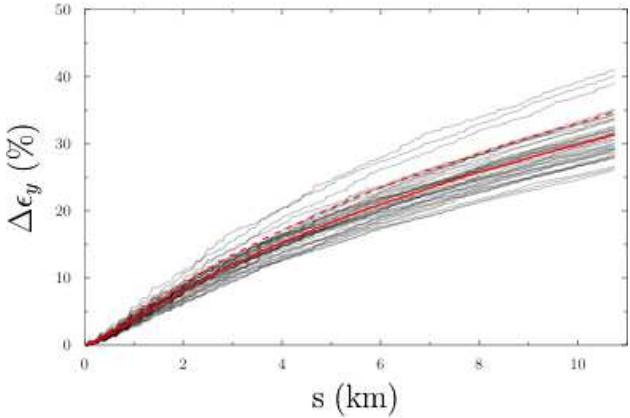

Figure 4: Emittance dilution simulated with Lucretia for 40 machines consisting of Ichiro cavities including the effects of random errors attributed to the manufacturing process. The mean curve is displayed in red, with one standard deviation dashed.

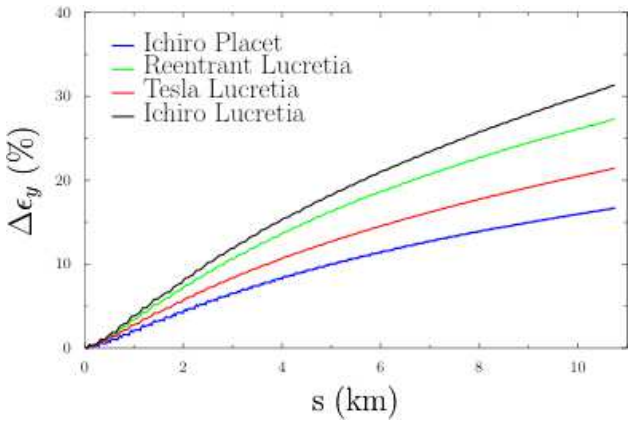

Figure 5: Comparison of the mean emittance dilution for each cavity type simulated with Lucretia. A single simulation undertaken with the code Placet for a linac comprising of Ichiro cavties is displayed in blue.

Earlier work has indicated that severe emittance dilution will occur should the machine be operating close to one of these resonances [15].

## CONCLUSIONS

Beam dynamics simulations have indicated that, with damping with Q~$10^6$, emittance dilution due to long range wakes in the Ichiro and Reentrant cavities is less than 50% at the end of the linac. Increasing This is tolerable performance similar to that of a linac made up of the ILC baseline TESLA cavities.

These simulations assume a bunch spacing of 369ns as detailed in the ILC Reference Design Report however, small deviations in this bunch spacing or equivalent systematic shifts in the HOM frequencies can lead to resonant wakefield effects, as illustrated in Fig. 6.

The damping required to preserve a high quality beam will be investigated with beam dynamics work including different levels of random detuning and systematic shifts to the HOM.

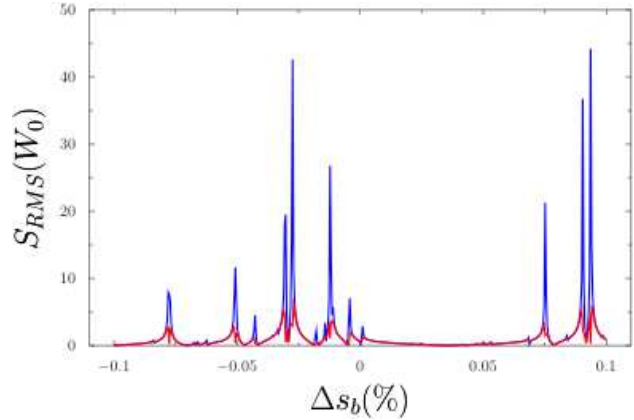

Figure 6: RMS of the sum wakefield as a function of fractional deviations in the bunch spacing. For Q=$10^6$ (blue) and Q=$10^5$ (red).

We note the importance of understanding the inconsistency between the Lucretia and Placet simulations. In order to shed light on the present discrepancy we intend to modify the mode randomization scheme to be consistent between both codes. Additional research is in progress on simulations of emittance dilution issues using Lucretia and including the short range wakefield due to cavity and coupler geometries and realistic component misalignments.